\documentclass{elsart}
\usepackage{amsmath,epsfig}
\usepackage{doublespace}
\begin{document}
\begin{frontmatter}
\title{Nonlinear molecular excitations in a completely  inhomogeneous DNA chain }
\author{M.~Daniel\corauthref{cor1}},
\ead{daniel@cnld.bdu.ac.in} 
\author{V.~Vasumathi}
\corauth[cor1]{Corresponding Author. Telephone:+91-431-2407057,  Fax:+91-431-2407093}
\address{ Centre for Nonlinear Dynamics, School of Physics, 
Bharathidasan University, Tiruchirappalli - 620 024, India.}
\date{} 
\begin{abstract}
   We study the nonlinear dynamics of a completely  inhomogeneous DNA  chain which is
     governed by a  perturbed sine-Gordon equation.  A multiple scale perturbation analysis provides perturbed kink-antikink solitons to represent open state configuration with small fluctuation. The perturbation due to inhomogeneities changes the velocity of the soliton. However, the width of the soliton remains constant.
\end{abstract}

\end{frontmatter}

\section{Introduction}
A number of theoretical models have been proposed in the recent times to study the nonlinear dynamics
     of Deoxyribonucleic acid (DNA)  molecule to understand the
 conservation and transformation of genetic information  (see for e.g \cite{ref1,ref2}). These models are based on
     longitudinal, transverse, rotational and  stretching  motions of bases.   Among these
  different  possible motions,
      rotational motion of bases  is found to contribute more towards the opening of base
      pairs and to the nonlinear dynamics of DNA. The first contribution towards nonlinear dynamics of DNA was made by Englander and his
     co-workers \cite{ref3} and they studied the base pair opening in DNA by taking into account the rotational motion. Yomosa \cite{ref4,ref5} proposed a plane base rotator model by taking into account the rotational motion of bases in a plane normal to the helical axis, and Takeno and Homma generalized the same \cite{ref6,ref7,ref8}. Later using this model, several authors found solitons to govern the fluctuation of DNA double helix between an open state and its equilibrium states \cite{ref11,ref12,ref13,ref14,ref15,ref16}. Peyrard and Bishop \cite{ref9,ref17} and Christiansen and his collegues \cite{ref10} studied the process of base pair opening by taking into account the transverse and longitudinal motions of bases in DNA.  Very recently, there have been extensions of the radial model of Bishop and Peyrard \cite{refpb1,refpb2}, composite models for DNA torsion dynamics \cite{refcom} and models interplaying between radial and torsional dynamics \cite{refrt1,refrt2,refrt3,refrt4}.
		 In all
	the above studies, homogeneous strands  and hydrogen bonds have been considered for the
	analysis.\\
~~~\\
      However, in nature the presence of different sites along the strands such as
	promotor, coding, terminator, etc., each of which has a  specific  sequence of bases is related
	to a particular function and thus making the strands
             site-dependent or inhomogeneous \cite{ref18,ref19}. Also, the presence of abasic sites leads to inhomogeneity  in stacking \cite{ref20}. In this context, in a  recent paper  the present authors \cite{ref21}  studied the nonlinear  molecular excitations in DNA with
  site-dependent stacking energy along the strands based on the plane base rotator model. The nonlinear dynamics of DNA in this case was found to be governed by a perturbed sine-Gordon (s-G) equation. The perturbed kink and antikink soliton solutions of the perturbed s-G equation represented an open state configuration of base pairs with small fluctuation.  The perturbation in this case introduces small fluctuations in the localized region of the soliton retaining the overall shape of the soliton. However, the width of the soliton remains constant and the velocity changes for different inhomogeneities.
 The results indicate that the  presence of inhomogeneity  in stacking changes the number of base pairs that participate in the open state configuration and modifies the speed with which the open state configuration  travels along the double helical chain. In reality, the presence of
  site-dependent strands in DNA  changes the nature of hydrogen bonds between adjacent base pairs and the  presence of abasic sites  leads to absence of hydrogen bonds. Thus, when the strands are site-dependent in stacking, naturally the hydrogen bonds that connect the bases between the strands  are also site-dependent. Hence, it has become necessary to consider inhomogeneity in hydrogen bonds  also  in the   study of nonlinear dynamics of DNA. In the present paper, we study the 
	dynamics of DNA  with  inhomogeneity  both in stacking and in hydrogen
	bonds using the plane
     base rotator model.  The paper is organized as follows. In section 2, we present the Hamiltonian of our model and derive the associated dynamical equation for the inhomogeneous DNA.  
       	 The effect of  stacking   and hydrogen bond inhomogeneity on base-pair opening is studied by solving the dynamical equations using a
	 multiple scale  soliton perturbation theory in section 3. The results are concluded in section 4.	 
 
\section{ Hamiltonian and  the dynamical equation}
We consider the B-form of a DNA double helix with site-dependent strands as well as base-pair sequence and study the nonlinear molecular excitations by considering a plane-base rotator model. In Fig. (1a) we   have presented a sketch of the DNA double helix. Here, $S$ and $S'$  represent the two complementary strands in the DNA double helix and each arrow  represents the direction of the base attached
     to the strand and the dots between arrows represent  the net hydrogen bonding effect between the 
     complementary bases. The z-axis is chosen along the helical axis of the DNA.     
      Fig. (1b) represents a horizontal projection of the $n^{th}$ base pair in the xy-plane. In this figure
      $Q_n$  and $Q'_n$ denote the  tips of the $n^{th}$ bases belonging to  the strands $S$ and $S'$. $P_n$ and $P'_{n}$ represent the points where the bases in  the $n^{th}$ base 
     pair are attached to the strands $S$ and
     $S'$ respectively.

  As we are looking for opening of base pairs in DNA which is related to  important DNA functions such as replication and transcription, we consider  the rotational motion of bases (due to its importance among other motions) in a plane normal to the helical axis (z-direction) represented by the angles $\phi_n$ and $\phi'_n$ at the $n^{th}$ site of the base pair. The stacking and hydrogen bonding energies are the major components of the energy in a DNA double helix. In the case of a homogeneous DNA system, Yomosa \cite{ref4,ref5} expressed the Hamiltonian involving these energies in terms of the rotational angles $\phi_n$ and $\phi'_n$ under the plane base rotator model which was later modified by the present authors \cite{ref21} in the case of site-dependent stacking. When both the stacking and hydrogen bonds are site-dependent, the  Hamiltonian for our  plane base rotator model of DNA double helix    is written in terms of the rotational angles as 
 \begin{eqnarray}
 H&=&\sum_n\left[ \frac{I}{2} ( {\dot\phi_n}^{2}+ {\dot\phi_n}^{'2}) +J f_n
 \left[2-\cos (\phi_{n+1}-\phi_n)
 -\cos  (\phi'_{n+1}-\phi'_n)\right]\right.\nonumber\\
&&\left. -\eta g_n\left[1-\cos  (\phi_n-\phi'_n)\right]\right].\label{eq1}
 \end{eqnarray}
    The first two terms in the Hamiltonian (\ref{eq1}) represent the kinetic energies of the rotational motion
  of the $n^{th}$  nucleotide bases with $I$    their moments of inertia and the remaining terms represent the potential energy due to stacking  and hydrogen bonds. While $J$ and $\eta$ represent a
   measure of stacking  and  hydrogen bonding
    energies respectively, $ f_n $ and $g_n$ indicate the site-dependent(inhomogeneous) character of 
    stacking  and hydrogen bonds respectively.\\
      The   Hamilton's equations of motion  corresponding to  
      Hamiltonian (\ref{eq1}) is   written as
     \begin{subequations}
  \begin{eqnarray}
  I\frac{\partial^2\phi_n}{\partial t^2}&=&J \left[f_n \sin (\phi_{n+1}-\phi_n)-f_{n-1}\sin
  (\phi_n-\phi_{n-1})\right]\nonumber\\
  &&+\eta g_n \sin (\phi_n-\phi'_n),
  \label{eq2a}\\
  I\frac{\partial^2\phi'_n}{\partial t^2}&=&J \left[f_n \sin (\phi'_{n+1}-\phi'_n)-f_{n-1}\sin
  (\phi'_n-\phi'_{n-1})\right]\nonumber\\
  &&+\eta g_n\sin (\phi'_n-\phi_n).\label{eq2b}
  \end{eqnarray}  
  \end{subequations}
It is expected that the difference in the angular rotation of neighbouring bases along the two strands in the case of B-form of DNA double helix   is  small \cite{ref6,ref7}. Therefore under small angle approximation, in the continuum limit Eqs. (\ref{eq2a}) and (\ref{eq2b}) are written as
  \begin{subequations}
\begin{eqnarray}
\phi_{\hat t\hat t}&=&f(z)\phi_{zz}+f_z \phi_{z} 
-\frac{1}{2} g(z)\sin (\phi-\phi'),\label{eq5a}\\
\phi'_{\hat t\hat t}&=&f(z)\phi'_{zz}+f_z
\phi'_{z}-\frac{1}{2}g(z) \sin (\phi'-\phi).\label{eq5b}
\end{eqnarray} 
\end{subequations} 
While writing Eqs. (\ref{eq5a}) and (\ref{eq5b}) we have  chosen $\eta=-\frac{Ja^{2}}{2}$  and rescaled the time variable  as $\hat{t}=\sqrt{\frac{Ja^{2}}{I}} t
$.  Now, adding and subtracting Eqs.(\ref{eq5a}) and (\ref{eq5b}) and by choosing
 $\phi=-\phi'$, we obtain 
\begin{eqnarray}
\Psi_{\hat t\hat t}-\Psi_{zz}+\sin\Psi=\epsilon \left[A\left( s(z)\Psi_{z}\right)_{z}- B~h(z)
\sin\Psi\right],\label{eq6}
\end{eqnarray}
where  $\Psi=2\phi$. Further, as the contribution due to inhomogeneity is small compared to homogeneous stacking and hydrogen bonding energies while writing Eq.(\ref{eq6}),  the inhomogeneity
 in stacking and hydrogen bonding energies are expressed in terms  of a small parameter $\epsilon$ as
$f(z)=1+\epsilon~A s(z)$ and $g(z)=1+\epsilon~B h(z)$, where $A$ and $B$ are  arbitrary constants.  When $\epsilon=0,$ Eq. (\ref{eq6}) reduces to the completely integrable sine-Gordon(s-G) equation   which admits N-soliton solutions in the form of kink and antikink \cite{ref22}. Hence, we call  Eq. (\ref{eq6}) as a
 perturbed sine-Gordon equation. For instance, the one soliton solution of the integrable s-G equation obtained through Inverse Scattering Transform (IST) method is written as,
\begin{eqnarray}
 \Psi(z,\hat t) =4 arctan{exp[\pm m_0(z-v_0\hat t)]} ,\quad m_0=\frac{1}{\sqrt{1-v_0^2}}.\label{eq7}
\end{eqnarray}
  Here $v_0$ and $m^{-1}_0$ are real parameters that represent the  velocity and width of the soliton 
respectively. In Eq.(\ref{eq7}) while the upper sign corresponds to the kink soliton, the lower sign represents the antikink soliton. In Figs. (2a) and (2b) the one soliton kink-antikink solutions (Eq.(\ref{eq7})) are plotted by choosing $v_0=0.4$. The kink-antikink soliton solution of the integrable sine-Gordon equation describes an open  state  configuration in the DNA double helix. In  Figure (2c), we present a sketch of how  the base
        pairs  open locally in the form of kink-antikink 
	structure in each strand and  propagate along the direction of the helical
	 axis. 

The base pair opening will help in the process of replication that duplicates DNA and in transcription which helps to synthesize messenger RNA. However, when $\epsilon\neq 0$, the inhomogeneity in stacking and hydrogen bonds may affect the base-pair opening through a perturbation on the kink-antikink solitons of the s-G equation. Therefore in the next section, we solve the perturbed s-G equation (\ref{eq6}) using a multiple-scale soliton perturbation theory \cite{ref32,ref33,ref34} (as has been carried out in the case  corresponding to $B=0$ \cite{ref21}) to understand the effect of stacking and hydrogen bond inhomogeneities on the base pair opening.  

\section{Effect of  stacking and hydrogen bonding inhomogeneities on the open state configuration} 
 \subsection{Multiple-scale soliton perturbation theory}
In order to study the effect of inhomogeneity in stacking and hydrogen bonds on the base pair opening in the form of kink-antikink soliton by treating them as a perturbation, the time variable $\hat t$ is transformed
into several variables as $t_n=\epsilon^{n} \hat t$ where $n=0, 1, 2,...$ and $\epsilon$ is a very small parameter \cite{ref32,ref33,ref34}.
 In view of this, the
time derivative  and $\Psi$ in Eq.~(\ref{eq6}) are replaced by the expansions
$
\frac{\partial }{\partial \hat t}=\frac{\partial }{\partial t_0}+\epsilon~
\frac{\partial }{\partial t_1}+\epsilon^2 \frac{\partial }{\partial
t_2}+...$ and $\Psi=\Psi^{(0)}+\epsilon \Psi^{(1)}+\epsilon^2 \Psi^{(2)}+....$
  We then equate
the coefficients of different powers of $\epsilon$ and obtain the following equations. At $O(\epsilon^{(0)})$ we obtain the  integrable
sine-Gordon equation   
\begin{eqnarray}
 \Psi^{(0)}_{t_0t_0}-\Psi^{(0)}_{zz}+sin\Psi^{(0)}=0, \label{eq8}
\end{eqnarray}
for which the one soliton solution takes the form  as given in Eq.(\ref{eq7}) with $\hat t$  replaced by $t_0$. Due to perturbation,
 the soliton parameters  namely $m$ and $\xi (\xi=v \hat t)$ are now treated
as functions of the slow time variables $t_0,t_1,t_2, .... $ However, $m$ is treated as independent
of $t_0$. 
 The equation at   $ O(\epsilon^{(1)})$  takes the form
\begin{eqnarray}
 {\Psi^{(1)}_{\tau\zeta}}-{\Psi_{\zeta\zeta}^{(1)}}+(1-2{sech^{2}}\zeta)\Psi^{(1)}=F^{(1)}
 (\zeta,\tau),\label{eq9}\\
 where\qquad\qquad\qquad\qquad\qquad\qquad\qquad\qquad\qquad\qquad\qquad\qquad\qquad\nonumber\\
  F^{(1)}(\zeta,t_{0})=2\left[As(\zeta) sech\zeta\right]_{\zeta}- 2aBh(\zeta)\tanh\zeta\mbox{sech}\zeta\qquad\qquad\nonumber\\
+4v_{0} sech\zeta\left[m_{t_{1}}+(m^{2}\xi_{t_{1}}-\zeta
m_{t_{1}})\tanh\zeta
 \right]. \label{eq10}
  \end{eqnarray} 
While writing the above equation we have used  the transformation
  $\hat \zeta=m(z-v t_{0})$ and $ \hat t_{0}=t_{0}$ to represent everything in a co-ordinate
  frame  moving with the soliton. Then, we have used  another set of
  transformations given by $\tau=\frac{\hat t_0}{2m}-\frac{(1+v)\hat \zeta}{2}$ and $\zeta=\hat \zeta$ for our later convenience.  We have also replaced $sin\Psi^{(0)}$ by $2a\tanh\zeta$ $sech\zeta$, where $a=\pm 1$. The solution of Eq.~(\ref{eq9}) is searched by assuming 
$\Psi^{(1)} (\zeta,\tau)=X(\zeta)T(\tau)$ and $
F^{(1)} (\zeta,\tau)= X_{\zeta}(\zeta)H(\tau).$
Substituting the above in Eq.~(\ref{eq9})  and simplifying we obtain 
\begin{eqnarray}
X_{\zeta\zeta}+(2 sech^{2}\zeta-1)X=\lambda_{0} X_{\zeta}, \label{eq11}\\
T_{\tau}-\lambda_{0} T= H(\tau),\label{eq12}
\end{eqnarray}
 where $\lambda_0$ is a constant.
 Thus, the problem of constructing the
perturbed soliton at this moment turns out to be solving 
 Eqs.~(\ref{eq11}) and (\ref{eq12}) by constructing the eigenfunctions and 
finding the eigen values \cite{ref21,ref34}. 
\\
It may be noted that Eq.(\ref{eq11})  differs from the normal eigen value problem, with $X_{\zeta}$ in the right hand side instead of $X$. Hence, before  actually solving the eigen value equation (\ref{eq11}), we  first consider it in a more general form   given by
 \begin{eqnarray}
 L_{1} X=\lambda \tilde{X} ,\qquad L_{1}=\partial_{\zeta\zeta}+ 2
sech^{2}\zeta -1, \label{eq14}
\end{eqnarray}
where $\lambda$ is the eigen value. To find the adjoint eigen function to $X$, we consider another eigen value problem

\begin{eqnarray}
 L_{2} \tilde{X}=\lambda X,\label{eq15}
 \end{eqnarray}
where $L_2$ is to be determined. Combining the two eigen value problems we get
\begin{eqnarray}
L_2 L_1 X= \lambda^2 X,~~
L_1 L_2 \tilde X= \lambda^2 \tilde X.\label{eq16}
\end{eqnarray}
From the above  equations we conclude that the operator $L_1 L_2$ is the adjoint of $L_2 L_1$ and also
 $X$ and $\tilde X$ are expected to be adjoint eigen functions.  Hence, we can find the eigenfunction by solving  Eqs.(\ref{eq14}) and (\ref{eq15}).  However, eventhough  $L_1$ is known as given in Eq.(\ref{eq14}), the operator  $L_2$ is still unknown. So, by experience we choose 
$ L_{2}=\partial_{\zeta\zeta}+6
sech^{2}\zeta -1$, and
  solve the eigen value equations  by choosing the eigen functions as
\begin{eqnarray}
X(\zeta,k)=p(\zeta,k) e^{ik\zeta},~~
\tilde {X}(\zeta,k)=q(\zeta,k) e^{ik\zeta}, \label{eq17}
\end{eqnarray}
where $k$ is the propagation constant.  Further, as the operator $L_2 L_1$  is self-adjoint, non-negative and satisfying a regular eigenvalue problem, the sine-Gordon soliton is expected to be stable. On substituting Eqs. (\ref{eq17}) in  Eqs.(\ref{eq14}) and (\ref{eq15}) in the asymptotic limit, we obtain the eigen value as $\lambda=-(1+k^{2})$.
Now, in order to find the eigen functions we expand  $p(\zeta,k)$ and $q(\zeta,k)$  \cite{ref32,ref34}  as
\begin{subequations}
\begin{eqnarray}
p (\zeta,k)&=&p_0+p_1 \frac{\sinh\zeta}{\cosh\zeta}+p_2 \frac{1}{\cosh^{2}\zeta}
+p_3
\frac{\sinh\zeta}{\cosh^{3}\zeta} 
+p_4\frac{1}{\cosh^{4}\zeta}+...,\label{eq19a}\\
q (\zeta,k)&=&q_0+q_1\frac{\sinh\zeta}{\cosh\zeta}+q_2 \frac{1}{\cosh^{2}\zeta}
+q_3
\frac{\sinh\zeta}{\cosh^{3}\zeta} 
+q_4\frac{1}{\cosh^{4}\zeta}+...,\label{eq19b}
\end{eqnarray}
\end{subequations}
where $p_j$ and $q_{j}$, j=0,1,2,... are functions of $k$ to be determined. On substituting Eqs. (\ref{eq17}), (\ref{eq19a}) and (\ref{eq19b}) in Eqs. (\ref{eq14}) and (\ref{eq15}) and collecting the coefficients of $ 1,~
\frac{\sinh\zeta}{\cosh\zeta}$,~  $\frac{1}{\cosh^{2}\zeta}$,... we get a set of simultaneous equations. On  solving those equations we obtain the following eigen functions \cite{ref21,ref34}.
\begin{eqnarray}
X(\zeta,k)&=&\frac{(1-k^{2}-2ik\tanh\zeta)}{\sqrt{2\pi}(1+k^{2})}e^{ik\zeta},
\label{eq20}\\
\tilde{X} (\zeta,k)&=&
\frac{(1-k^{2}-2ik\tanh\zeta-2sech^{2}\zeta)}{\sqrt{2\pi}(1+k^{2})} e^{ik\zeta}.
 \label{eq21}
\end{eqnarray}
The higher order coefficients $p_3, p_4,...$ and $q_3, q_4,...$ vanish.\\
It may be noted that  Eq.(\ref{eq12})   is a linear inhomogeneous differential equation and  can be solved using known procedures. The solution reads
\begin{eqnarray}
  T(\tau,k)=\frac{1}{i\lambda_{0} k(1+k^{2})}\int_{-\infty}^{\infty}
d\zeta'  F^{(1)} (\zeta',\tau){X}^{\ast}(\zeta',k)
(e^{\lambda_{0}
[\tau-\frac{(1+v)}{2}\zeta']}-1),\label{eq13}
\end{eqnarray}
where $\lambda_{0}=\frac{i(1+k^{2})}{k}$.
 The first order correction to the soliton can be computed  using the following expression.
 \begin{eqnarray}
\Psi^{(1)}(\zeta,\tau)=\int_{-\infty}^{\infty} X(\zeta,k) T(\tau,k)
dk+\sum_{j=0,1}X_{j} (\zeta) T_{j}(\tau).\label{eq22}
\end{eqnarray}
Here the continuous eigenfunctions $X(\zeta,k)$ and $ T(\tau,k)$ are already known as given in Eqs. (\ref{eq20}) and (\ref{eq13}). However, the discrete eigen states  $X_{0}, X_{1}$ and $T_{0},T_{1}$ are unknown.  The  two   discrete eigenstates $X_{0}, X_{1}$  corresponding to the discrete eigen value $\lambda=0$   can be found   using the completeness  of the continuous eigenfunctions as
\begin{eqnarray}
 X_{0} (\zeta)&=& sech\zeta ,~X_{1} (\zeta)= \zeta sech\zeta.\label{eq23}
\end{eqnarray}

In order to find $T_{0}$ and $T_{1}$, we
 substitute Eq.(\ref{eq22}) in Eq.(\ref{eq9}) and multiply by $X_0(\zeta)$ and $X_1(\zeta)$ separately and use the orthonormal relations to get
\begin{subequations}
\begin{eqnarray}  
{ T_{1}}_{\tau} (\tau)&=&\int_{-\infty}^{\infty} F^{(1)} (\zeta,\tau) X_{0} (\zeta)
d\zeta , \label{eq24a}\\
 {T_{0}}_{\tau} (\tau)-2T_{1} (\tau)&=&-\int_{-\infty}^{\infty} F^{(1)}
 (\zeta,\tau)
X_{1} (\zeta) d\zeta.  \qquad\label{eq24b}
\end{eqnarray}
\end{subequations}
As $F^{(1)}(\zeta,\tau)$  given in Eq.(\ref{eq10}) does not contain the time  variable $\tau$ explicitly, the right hand side of Eqs. (\ref{eq24a}) and (\ref{eq24b})  also should be  independent of time, and hence  we write  the nonsecularity conditions  as
\begin{subequations}
\begin{eqnarray}
\int_{-\infty}^{\infty}F^{(1)} (\zeta,\tau) X_{0} (\zeta) d\zeta=0, \label{eq25a}\\
\int_{-\infty}^{\infty}F^{(1)} (\zeta,\tau) X_{1} (\zeta) d\zeta=0. \label{eq25b} 
\end{eqnarray}
\end{subequations}
 It may be indicated that, the nonsecularity conditions give way for the stability of soliton \cite{refsh} under perturbation through an understanding of the evolution of soliton parameters such as width and velocity.
Substituting the above equations in Eqs. (\ref{eq24a}) and (\ref{eq24b}), we choose $T_1(\tau)=0$ and find $T_0(\tau)=C$  which has to be determined. For this, we substitute $T_1(\tau)=0$ in Eq. (\ref{eq24b}) and integrate to obtain
\begin{eqnarray} 
T_{0}(\tau)&=& \frac{(1+v)}{2}\int_{-\infty}^{\infty}d\zeta ~\zeta F^{(1)}(\zeta,\tau) X_{1}(\zeta)
.\label{eq26}
\end{eqnarray}
In order to find the first order correction  to soliton $\Psi^{(1)}$ we need to evaluate the eigen states   $T(\tau,k)$ and $T_0(\tau)$
  explicitly for  which we have to find the explicit form of  $F^{(1)}(\zeta,\tau)$ which contains unknown quantities like $m_{t_{1}},~\xi_{t_{1}},~ s(\zeta)$ and $h(\zeta)$.  Therefore, we substitute Eqs. (\ref{eq10}) and (\ref{eq23}) in  the nonsecularity conditions given in Eqs. (\ref{eq25a}) and (\ref{eq25b}) and evaluate the integrals to obtain 
\begin{subequations}
\begin{eqnarray}
m_{t_{1}}&=&-\frac{1}{2v_0} \int_{-\infty}^{\infty} \left(A\left[ {s(\zeta) sech\zeta}\right]
_{\zeta} -a B h(\zeta)sech\zeta\tanh\zeta \right) sech\zeta d\zeta,  \label{eq27a}\\
\xi_{t_{1}}&=&-\frac{1}{2m^{2} v_0}\int_{-\infty}^{\infty} \left(A\left[ {s(\zeta) sech\zeta}\right]
_{\zeta}   - aB h(\zeta)sech\zeta\tanh\zeta\right) \zeta sech\zeta d\zeta .\label{eq27b}
\end{eqnarray}
\end{subequations}
 Here $m_{t_1}$ and $\xi_{t_1}$ represent the time variation of the inverse of the width and  velocity  of the soliton respectively. 
\subsection{ Variation of soliton parameters}
In order to find the variation of the width and velocity of the soliton while propagating along the inhomogeneous DNA chain, we assume that the soliton with a width of $m_0^{-1}$ is travelling with a speed of $v_0 (\xi_{t_0})$ when the perturbation is switched on. In other words,  the problem now boils down to understand the propagation of the kink-antikink soliton  in an inhomogeneous DNA chain and to measure the change of the soliton parameters due to inhomogeneity. To find the variation of the soliton parameters explicitly and to construct the perturbed soliton solution we have to evaluate the integrals found in  the right hand sides of Eqs. (\ref{eq27a}) and (\ref{eq27b}) which can be carried out only on supplying specific forms of $s(\zeta)$ and $h(\zeta)$.  Hence, we consider  inhomogeneity  in the form of  
localized and 
periodic functions separately and further assume that the inhomogeneity in the stacking and in the hydrogen bonds are equal. The localized inhomogeneity represents  the intercalation of a
compound between neighbouring base pairs or the presence of defect or the presence of abasic site in the 
DNA double helical chain. The periodic nature of inhomogeneity  may represent a
 periodic
repetition of similar base pairs along the helical chain. 
 
\subsubsection{ Localized inhomogeneity}  
To understand the effect of localized inhomogeneity in stacking and hydrogen bonds on the  width and velocity of the soliton during propagation, we substitute
 $s(\zeta)=h(\zeta)=sech\zeta$   in Eqs.(\ref{eq27a}) and (\ref{eq27b}). On evaluating the integrals, we obtain  $m_{t_{1}}=0, ~ \xi_{t_{1}}=\frac{\pi}{12m^{2}v_{0}}(2 A+aB)$ which can be written
in terms of the original time variable $\hat t$ by using the expansions $m_{\hat{t}}=m_{t_0}+\epsilon m_{t_1}$ and $\xi_{\hat{t}}=\xi_{t_0}+\epsilon \xi_{t_1}$ as
 \begin{eqnarray}
m=m_{0}, \quad \xi_{\hat{t}}\equiv v=v_{0}+\frac{\epsilon\pi(2
A+aB)}{12m^{2}v_{0}}, \label{eq28}
\end{eqnarray}
where $1/m_{0}$ is the  width  and $v_0$ is the  velocity of the soliton in  the absence of inhomogeneity.  The first  of Eq.(\ref{eq28}) says that when the inhomogeneities in stacking and  in hydrogen bonds are in the form $s(\zeta)=h(\zeta)=sech\zeta$, the width  of the soliton remains constant, thereby showing that the number of base pairs participating in the opening process remain constant during propagation. However, from the second of Eq.(\ref{eq28}), we find that the velocity of the soliton gets a correction.      The nature of correction in the velocity depends on the value of `$a$' which takes $+1$  or $-1$ and also on the nature of $A$ and $B$ which can be either positive or negative. When $A$ and $B$ are greater than zero, the inhomogeneity will correspond to an energetic barrier and on the other hand when $A$ and $B$ are less than zero, the inhomogeneity will correspond to $a$ potential well.  First, we consider the case corresponding to $a=+1$. In this case when $(2A+B)>0$, the velocity of the soliton gets a positive correction and it may propagate along the chain without formation of a bound state. On the other hand, when $(2A+B)<0$, the inhomogeneities slow down the soliton. Ofcourse, when $(2A+B)=0$, that is when the inhomogeneities in the stacking and in the hydrogen bonding suitably balance each other, the velocity of the soliton remains unaltered. Finally, the soliton stops when the original velocity satisfies the condition $v_0^2=\frac{1}{1-\frac{12}{\epsilon\pi(2A+B)}}$. In all the above cases, the stability of the soliton is guaranteed. A similar argument can be made in the case when $a=-1$ with $(2A+B)$ replaced by $(2A-B)$.  It may also be noted that similar results have been obtained in the case of resonant kink-impurity interaction and kink scattering in a perturbed sine-Gordon model by Zhang Fei  et al \cite{ref36},  say that the kink will pass the impurity and escape to the positive infinity when the initial velocity of kink soliton is larger than the critical value. In a similar context Yakushevich et al \cite{ref36a} while studying the interaction between soliton and the point defect in DNA chain showed numerically, that the solitons are stable.  At this point it is also worth mentioning that Dandoloff and Saxena \cite{ref35} realized that in the case of an XY-coupled spin chain model which is identifiable with our  plane-base rotator model of DNA, the ansatz $sech\zeta$ energetically favours the deformation of spin chain.\\
 \subsubsection{ Periodic inhomogeneity} 
We  then choose the  periodic inhomogeneity  in the form  $s(\zeta)=h(\zeta)=\cos\zeta$  and substitute the same in Eqs. (\ref{eq27a}) and (\ref{eq27b}). On  evaluating the integrals, we get $m_{t_{1}}=0, ~ \xi_{t_{1}}=\frac{\pi[\pi A+a(4-\pi)B]}{16m^{2}v_{0}}$ which can be written in terms of the original time variable $\hat t$ as
\begin{eqnarray}
m=m_{0}, \quad \xi_{\hat{t}}\equiv v=v_{0}+ \frac{\epsilon\pi[\pi A+a(4-\pi)B]}{16m^{2}v_{0}}.
\label{eq29}
\end{eqnarray}
From  Eq.(\ref{eq29}), we observe that  the  width of the soliton remains 
 constant and the velocity gets a correction.  Here also, one can project an argument similar to the case of localized inhomogeneity. The only difference between the two cases is the quantum of correction added to soliton velocity. Normally in DNAs, inhomogeneity in hydrogen bonds is expected to be dominant and therefore one would expect that $B>A$. One can verify that it is possible to obtain the above condition from the velocity corrections in Eqs. (\ref{eq28}) and (\ref{eq29}) by writing  $\frac{\epsilon\pi[\pi A+(4-\pi)B]}{16m^{2}v_{0}}>\frac{\epsilon\pi(2
A+B)}{12m^{2}v_{0}}$ when $a=+1$. The above condition indicates that the correction in velocity in the case of periodic inhomogeneity is larger than that in the case of localized inhomogeneity. This is because in this case, the inhomogeneity occurs periodically in the entire length of the DNA chain. In a recent paper, Yakushevich et al \cite{ref36a}  studied numerically the dynamics of topological solitons describing open states in an inhomogeneous DNA and investigated interaction of soliton with the inhomogeneity and the results, have very close analogy with the results of our perturbation analysis. It was shown that the soliton  can easily propagate along the DNA chain without forming a bound state thus showing that soliton moving with sufficiently large velocity along the DNA chain is stable with respect to defect or inhomogeneity.
\subsection{First order perturbed soliton}
Having understood the variation of the width and velocity of the soliton in a slow time scale due to perturbation, we now construct the first order perturbed soliton  by substituting the values of the basis functions $\{X\}
\equiv\{X(\zeta,k), X_{0}(\zeta), X_{1}(\zeta)\}$ and
 $\{T\}\equiv\{T(\tau,k), T_{0}(\tau), T_{1}(\tau)\}$  given in Eqs. (\ref{eq20}), (\ref{eq23}), (\ref{eq13}), (\ref{eq26}) with  $T_1(\tau)=0$ in Eq.(\ref{eq22}), we get
\begin{eqnarray}
\Psi^{(1)}(\zeta,t_{0})&=&-\frac{1}{\pi}\int_{-\infty}^{\infty} dk \frac{1}{(1+k^{2})^3}(1-k^{2}-2ik\tanh\zeta)  e^{ik\zeta}\nonumber\\
&&\times\int_{-\infty}^{\infty}
d\zeta' \left(\left[As(\zeta') sech\zeta' \right]_{\zeta'}-aBh(\zeta')sech\zeta'\tanh\zeta'\right.
+2v_0~sech\zeta'\nonumber\\
 &&\times\left.\left[m_{t_{1}}+(m^{2}\xi_{t_{1}}-\zeta'
 m_{t_{1}})\tanh\zeta' \right] \right)
  (1-k^{2}+2ik\tanh\zeta')  e^{-ik\zeta'}\nonumber\\
&& \times\left(e^{\lambda_{0}
[\tau+\frac{(1+v_{0})}{2}\zeta']}-1\right)+(1+v)sech\zeta\int_{-\infty}^{\infty}d\zeta'\left(\left[As(\zeta') sech\zeta' \right]_{\zeta'}\right.\nonumber\\
&&-aBh(\zeta')sech\zeta'\tanh\zeta'+2v_0~sech\zeta'\left[m_{t_{1}}\right.\nonumber\\
&&\left.+(m^{2}\xi_{t_{1}}-\zeta'
\left.m_{t_{1}})\tanh\zeta' \right]\right)\zeta'^{2} sech\zeta'.\label{eq30}
\end{eqnarray}
While writing the above, we have also used $ \tau=\frac{1}{2m}[t_{0}-m(1+v)\zeta]$.  It may be noted that majority of the poles that lie within the contour in Eq. (\ref{eq30}), are purely imaginary giving rise to exponentially localized residues and hence do not give rise to any radiation thus will lead to stable form of soliton \cite{refpb6}.
 
\subsubsection{Localized inhomogeneity}
 Now, we explicitly construct the first order perturbation correction to the one soliton in the case of the localized inhomogeneity (i.e) when $s(\zeta)=h(\zeta)=sech\zeta$,  by substituting the corresponding values 
  of $F^{(1)},  m_{t_{1}} $ and $\xi_{t_{1}}$  in Eq.~(\ref{eq30}). 
  The resultant integrals are  then evaluated using  standard residue theorem \cite{ref37}  which involves very lengthy algebra and at the end we  obtain  
\begin{eqnarray}
\Psi^{(1)}(\zeta,t_{0})&\approx&(2A+aB)\left[\frac{40}{27\sqrt{2}}\sqrt{sech\zeta}
~e^{-\frac{3\zeta}{2}}
 +\frac{32}{27\sqrt[3]{2}}\tanh\zeta
\sqrt[3]{sech\zeta}~e^{-\frac{5}{3}\zeta} \right.\nonumber\\
&& +\frac{\pi}{12v^{2}} 
\left.[2 v t_0+m^2(v^2-1)]sech\zeta\right].  \label{eq31}
\end{eqnarray}
Finally,  the perturbed one soliton solution,  that is
$\Psi(z,t_{0})=\Psi^{(0)}(z,t_{0})+
\Psi^{(1)}(z,t_{0})$ (choosing $\epsilon=1$) is written 
  in terms of the original variables as
\begin{eqnarray}
\Psi(z,t_{0})&\approx &4
arc\tan\exp[\pm m_{0}(z-v_{0}t_{0})]+(2A+aB)\left[\frac{40}{27\sqrt{2}}e^{\mp\frac{3(m(z-v t_{0})}{2}}\right.\nonumber\\
&&
\times\sqrt{sech[\pm m(z- v t_{0})]}+\frac{32}{27\sqrt[3]{2}}
\tanh[\pm m(z-v t_{0})]\nonumber\\ 
&& \times\sqrt[3]{sech[\pm m(z-v t_{0})]} 
 e^{\mp\frac{5}{3}m(z-v t_{0})}+\frac{\pi}{12 m v^{2}}\nonumber\\
 &&\left.\times 
\left[m (v^{2}-1)+2 t_{0} v\right]sech[\pm m(z-v t_{0})]\right].  \label{eq32}
\end{eqnarray}

Knowing $\Psi$, the angle of rotation of bases $\phi(z,t_{0})$  can be immediately written down by using the
   relation $\phi=\frac{\Psi}{2}$.
In Figs. 3(a,b)  we have  plotted  $\phi(z,t_{0})$, the rotation of bases under the perturbation  
  $g(z)=h(z)=sech z$  by choosing  $a=A=B=1$  and $v_{0}=0.4$. From the figures we observe that there appears fluctuation in the form of a train of pulses closely resembling the shape of the inhomogeneity profile in the width of the soliton as time progresses.  But, there is no change in the topological character and no fluctuations appear in the asymptotic region of the soliton.  In Figs. (4a) and (4b), we have plotted the perturbed kink and antikink solitons respectively corresponding to the case when the hydrogen bond inhomogeneity is absent (B=0) for comparison. From  Figs. (3) and (4), we observe that the fluctuation in the form of a train of pulses appear in  both the cases. Eventhough the perturbed solitons in both the cases appear qualitatively the same, the inhomogeneity in hydrogen bonds adds more fluctuation in the width of the soliton.
\subsubsection{Periodic inhomogeneity}
 We  then repeat the procedure  for constructing the perturbed one
soliton solution $\Psi(\zeta,t_{0})$ in the case of periodic inhomogeneity by choosing
  $s(\zeta)=h(\zeta)=\cos\zeta$, and    obtain  the perturbed soliton solution as
\begin{eqnarray}
 \Psi(z,t_{0})&\approx&4~
arctan\exp[\pm m_{0}(z-v_{0} t_{0})]
+[A \pi+aB(4-\pi)] [b+d t_{0}]\nonumber\\
&&sech[ m_0(z-v t_{0})],\label{eq33}
 \end{eqnarray}
where $b=\frac{\pi^{2}}{16  v^{2}}
 (v^{2}-1)$ and $d=\frac{\pi^{2}}{8 m_0 v}$.  In Figs. 5(a,b), we plot the angle of rotation of bases  $\phi$($=\frac{\Psi}{2}$), using the perturbed soliton given in Eq.(\ref{eq33}) for the same parametric choices as in the case of localized inhomogeneity. We observe from the figures that, the fluctuation appears in the width of the soliton without any change in its topological character asymptotically. In order to compare the results with that of the case when the hydrogen bonding inhomogeneity is absent as also found in  \cite{ref21}(see Figs. (6a,b)), we plot the perturbed soliton found in (\ref{eq33}) when $B=0$.  As in the previous case, here also we observe that fluctuation  appears in the width of the soliton in both the cases and the inhomogeneity in hydrogen bonds  introduces more fluctuation.
\section{Conclusions } 
 In this paper, we 
  studied the nonlinear dynamics of a completely inhomogeneous (inhomogeneity in both  stacking and hydrogen bonding)  DNA  double helix by considering the dynamical plane-base rotator model. The dynamics of this model in the continuum limit gives rise to a perturbed sine-Gordon equation, which was derived from the Hamiltonian consisting of site-dependent stacking and  hydrogen bonding energies. In the unperturbed limit, the dynamics is governed by the kink-antikink soliton of the integrable sine-Gordon equation which represents the opening of base pairs  in a homogeneous DNA. In order to understand the effect of inhomogeneity in stacking and hydrogen bonds on the base pair opening, we carried out a perturbation analysis  using multiple-scale soliton perturbation theory. The perturbation not only modifies the shape of the soliton but also introduces change in the velocity of the soliton. 
    From the results, we observe that when the inhomogeneity is in a localized or periodic form, the width of the soliton remains constant.  However, the velocity of the soliton increases, decreases or remains uniform and even the soliton stops, depending on the values of inhomogeneity represented by $A$ and $B$. The soliton in all the cases are found to be stable.  From the results of the perturbed soliton  we observe that the inhomogeneity in stacking and hydrogen bonds  in both the cases (localized and periodic forms) introduce fluctuation in the form of pulses in the width of the soliton.  However, there is no change in the topological character of the soliton in the asymptotic region(see Figs. 3 and 5). The fluctuation is more when both the inhomogeneities (site-dependent stacking and hydrogen bonds) are present, whereas it is less in the case of homogeneous hydrogen bonds and site-dependent stacking. Hence, we conclude that,  the addition of inhomogeneity in hydrogen bonds does not introduce big changes in the soliton parameters and   shape   except a correction in the velocity of the soliton and  fluctuation.  The above dynamical behaviour may act as energetic activators of the RNA-polymerase transport process during transcription in DNA.  In the case of short DNA chains the  discreteness effect assumes importance and hence we will analyse the discrete dynamical equations (2)  and the results will be published elsewhere.
\section{Acknowledgments}
   The work of M. D forms part of a major
  DST  project.  V. V  thanks SBI for financial support.
 
\newpage
Fig.1. (a) A schematic structure of the B-form  DNA double helix. (b) A horizontal projection of the $n^{th}$ base
pair in the xy-plane.\\
Fig.2. (a) Kink  and (b) antikink soliton solutions of the sine-Gordon equation (Eq.(\ref{eq6}) when
$\epsilon=0$ ) with $v_0=0.4$.
 (c) A sketch of the formation of open state configuration in terms of
kink-antikink soliton in DNA double helix.\\
Fig.3.(a) The perturbed kink-soliton and (b) the perturbed antikink-soliton for the inhomogeneity
$g(z)=h(z)=sech z$ with $A=B=1$ and $v_0=0.4$.\\
Fig.4. (a) The perturbed kink-soliton and (b) the perturbed antikink-soliton for the inhomogeneity
$s(z)=sech z,h(z)=0$ with $A=1$ and $v_0=0.4$.\\
Fig.5.(a) The perturbed kink-soliton and (b) the perturbed antikink-soliton for the inhomogeneity
$g(z)=h(z)=\cos z$ with $A=B=1$ and $v_0=0.4$.\\
Fig.6.(a) The perturbed kink-soliton and (b) the perturbed antikink-soliton for the inhomogeneity
$g(z)=\cos z,h(z)=0$ with $A=1$ and $v_0=0.4$.\\
\begin{figure}
\begin{center}
\psfig{file=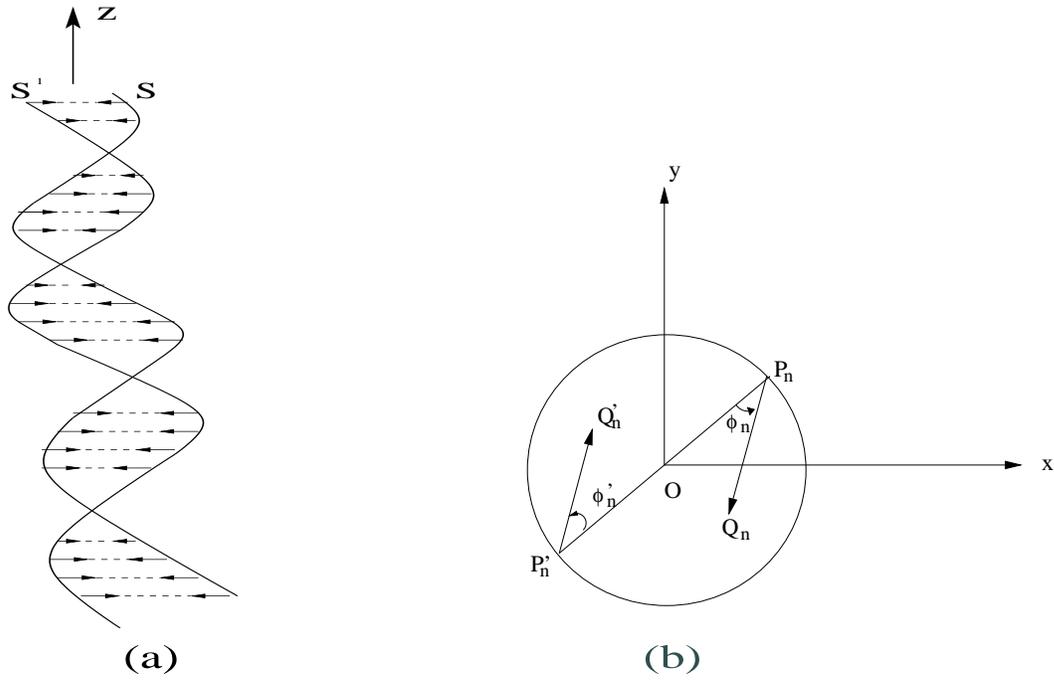,height =9cm, width=14cm}
\caption{(a) A schematic structure of the B-form  DNA double helix. (b) A horizontal projection of the $n^{th}$ base
pair in the xy-plane.}
\end{center}
\end{figure} 
\begin{figure}
\begin{center}
\epsfig{file=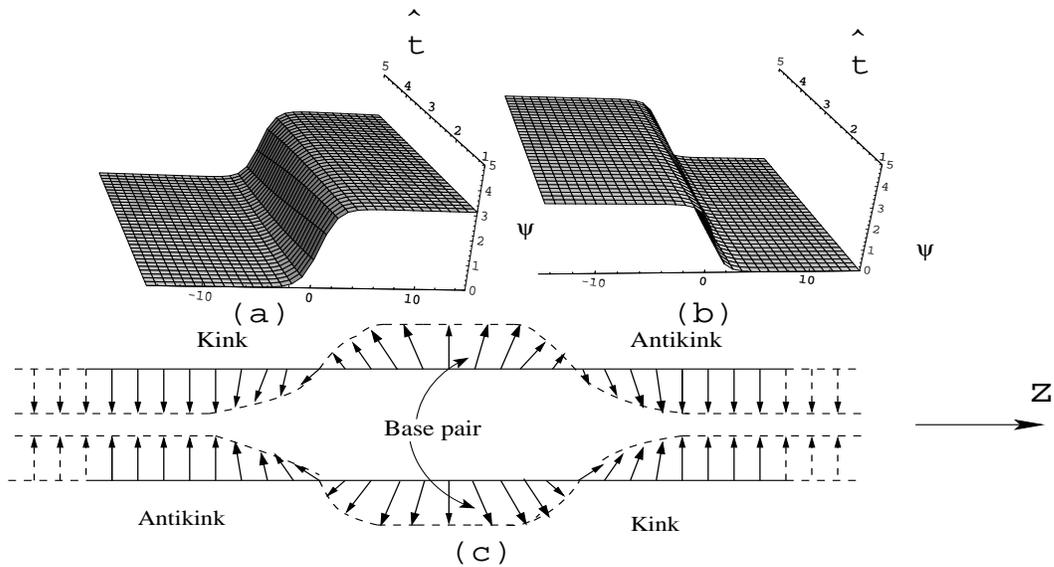,height =9cm, width=14cm}
\caption{(a) Kink  and (b) antikink soliton solutions of the sine-Gordon equation (Eq.(\ref{eq6}) when
$\epsilon=0$ ) with $v_0=0.4$.
 (c) A sketch of the formation of open state configuration in terms of
kink-antikink soliton in DNA double helix.}
\end{center}
\end{figure}
\begin{figure}
\begin{center}
\epsfig{file=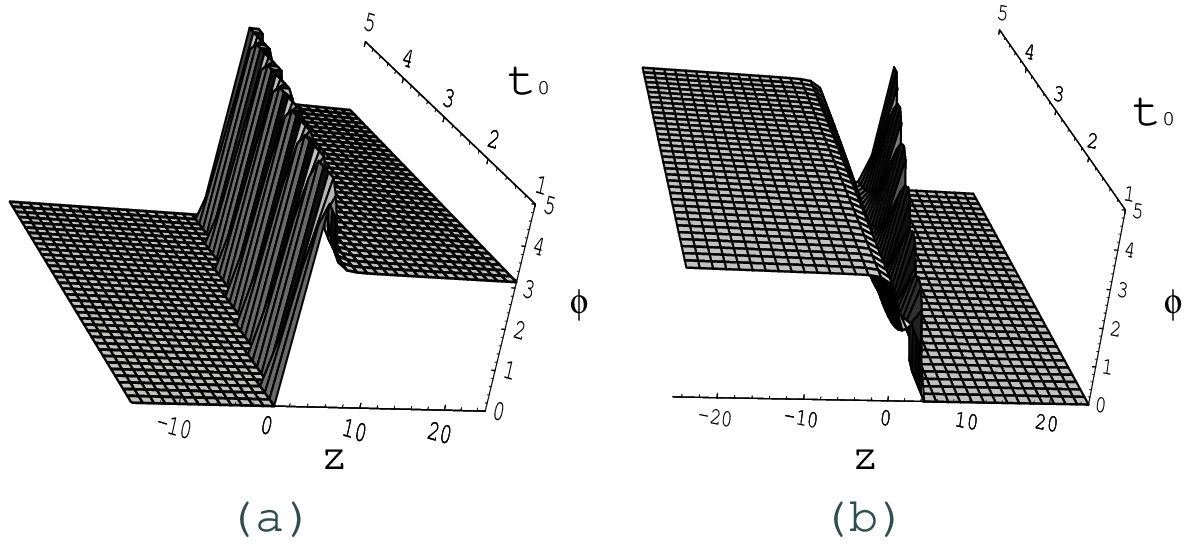,height =6cm, width=14cm}
\caption{(a) The perturbed kink-soliton and (b) the perturbed antikink-soliton for the inhomogeneity
$g(z)=h(z)=sech z$ with $A=B=1$ and $v_0=0.4$.}
\end{center}
\end{figure}
 \begin{figure}
\begin{center}
\epsfig{file=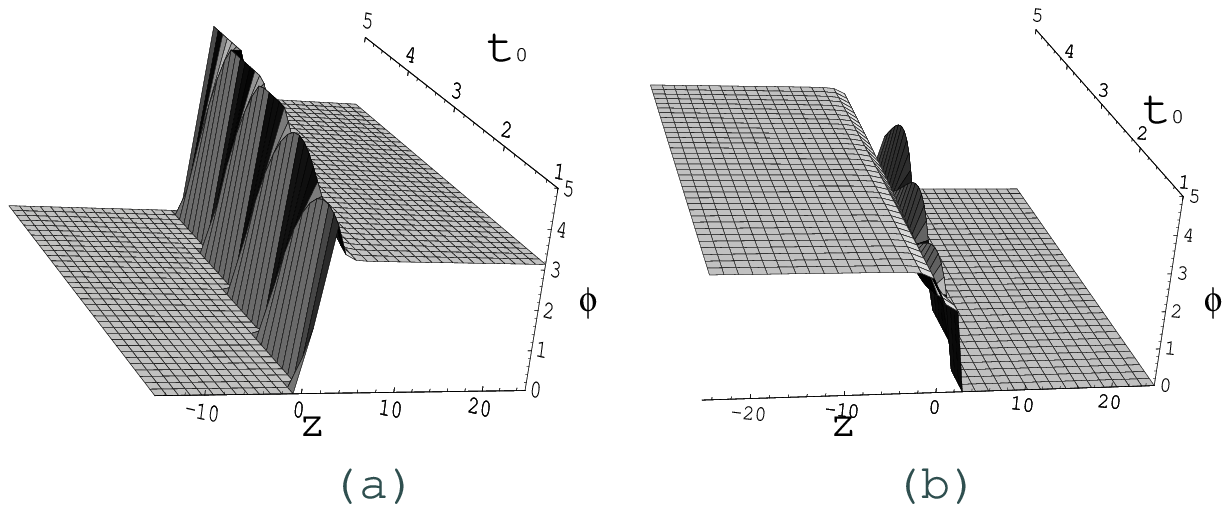,height =6cm, width=14cm}
\caption{(a) The perturbed kink-soliton and (b) the perturbed antikink-soliton for the inhomogeneity
$s(z)=sech z,h(z)=0$ with $A=1$ and $v_0=0.4$.}
\end{center}
\end{figure}
\begin{figure}
\begin{center}
\epsfig{file=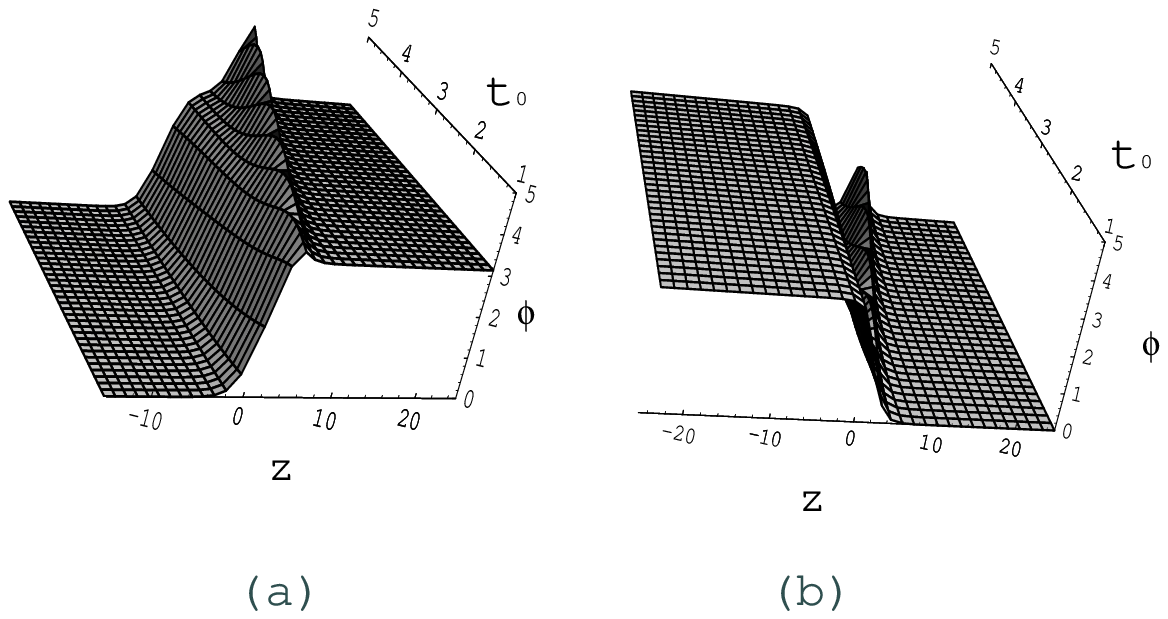,height =6cm, width=14cm}
\caption{(a) The perturbed kink-soliton and (b) the perturbed antikink-soliton for the inhomogeneity
$g(z)=h(z)=\cos z$ with $A=B=1$ and $v_0=0.4$.}
\end{center}
\end{figure}
 
\begin{figure}
\begin{center}
\epsfig{file=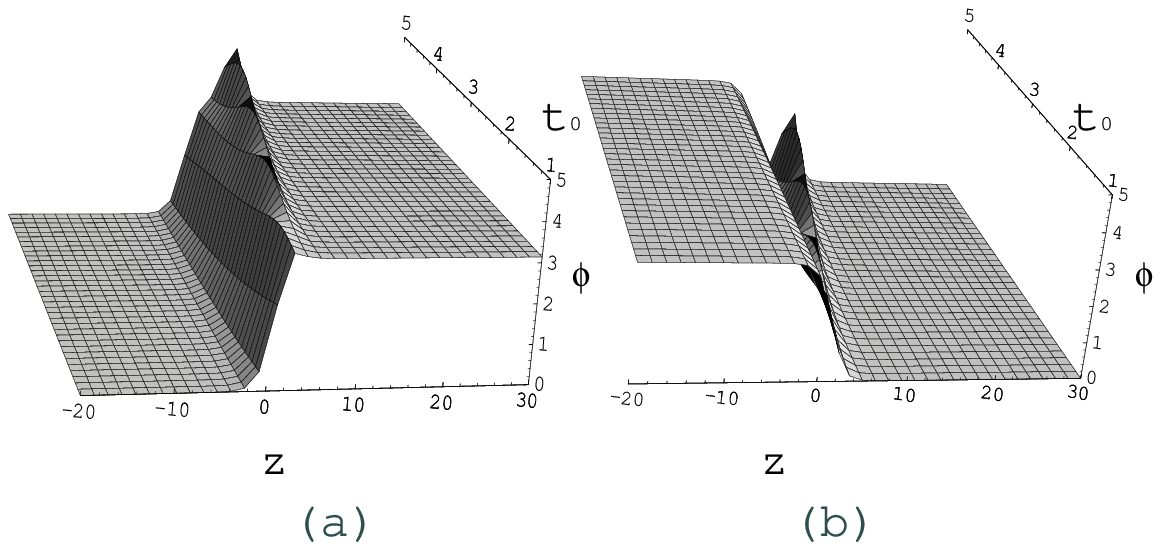,height =6cm, width=14cm}
\caption{(a) The perturbed kink-soliton and (b) the perturbed antikink-soliton for the inhomogeneity
$g(z)=\cos z,h(z)=0$ with $A=1$ and $v_0=0.4$.}
\end{center}
\end{figure}
\end{document}